# An Event-Based Synchronization Framework for Controller Hardware-in-the-loop Simulation of Electric Railway Power Electronics Systems

Jialin Zheng, *Student Member, IEEE*, Yangbin Zeng, *Member, IEEE*, Han Xu, *Student Member* IEEE, Weicheng Liu, *Student Member, IEEE* and Di Mou, *Member, IEEE*, Zhengming Zhao, *Fellow, IEEE*,

***Abstract*- The Controller Hardware-in-the-loop (CHIL) simulation is gaining popularity as a cost-effective, efficient, and reliable tool in the design and development process of fast-growing electrified transportation power converters. However, it is challenging to implement the conventional CHIL simulations on the railway power converters with complex topologies and high switching frequencies due to strict real-time constraints. Therefore, this paper proposes an event-based synchronization CHIL (ES-CHIL) framework for high-fidelity simulation of these electrified railway power converters. Different from conventional CHIL simulations synchronized through the time axis, the ES-CHIL framework is synchronized through the event axis. Therefore, it can ease the real-time constraint and broaden the upper bound on the system size and switching frequency. Besides, models and algorithms with higher accuracy, such as the diode model with natural commutation processes, can be used in the ES-CHIL framework. The proposed framework is validated for a 350 kW wireless power transformer system containing 24 fully controlled devices and 36 diodes by comparing it with Simulink® and physical experiments. This research improves the fidelity and application range of the power converter's CHIL simulation. Thus, it helps to accelerate the prototype design and performance evaluation process for electrified railways and other applications with such complex converters.** [1]

***Index Terms*—Electric railway, Wireless power transfer (WPT), Controller Hardware-in-the-loop (CHIL).**

## I. Introduction

The share of rail transport in total transport activity need to reach at least 13% by 2030 to achieve the net zero aims, thus requiring more efficient technologies, such as the wireless power transfer (WPT) converters in Fig 1, etc., to support a rapid and widespread transition to rail electrification [1]. During the electrification transition, there is an urgent need for rapid and reliable analysis and evaluation of key power converters in electrified rail transportation [2].

The controller hardware-in-the-loop (CHIL) simulation is gaining popularity as an effective and reliable analysis and evaluation tool, due to its ability to avoid iterative testing of expensive and time-consuming hardware prototypes, and significantly improve safety and reduce costs [3]–[5]. However, with the application of wide bandgap power semiconductor devices and new power electronic topologies, electrified railway power converters are becoming more complex and have higher switching frequencies [6], [7]. Due to these characteristics, the CHIL simulation of these railway power converters is still difficult and faces the following three main challenges.

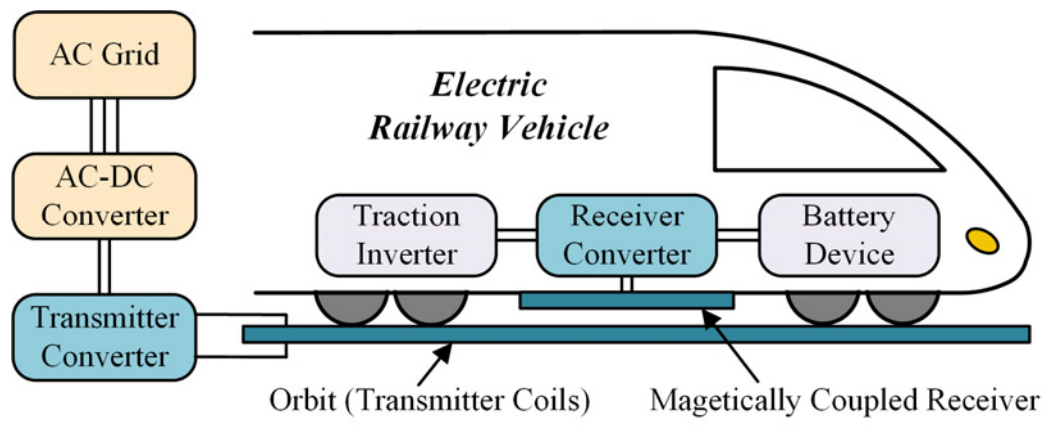

**Fig. 1** Schematic diagram of wireless power transport for electric railway.

**1)** ***Topology complexity***: The power converters of electrified railway generally use a modular combination of power supplies to meet power demands that can reach hundreds of kW to MW power levels [8]. The problem is that the increased system circuit size leads to the increased computational cost of solving the circuit equations in CHIL simulations, which is exacerbated by the time-varying topology due to power switching states [9].

Several approaches have been proposed in modeling and solving to reduce computational cost. For modeling, the associated discrete circuit (ADC) model is used to ensure that the system conduction matrix no longer varies with the power switching state [10], but it poses the problem of virtual loss [11]. Several improved ADC models are proposed for reducing the virtual loss problem [11], [12]. For solving, a matrix decoupling method is proposed to tear the system matrix by multiple solvers, which requires a lot of effort in large systems applications [13]. A latency insertion method (LIM) is proposed to achieve branch-level decoupling by inserting small inductors and capacitors to reduce the computational cost, but the inserted inductors and capacitors will change the dynamic characteristics of the whole system [14]. In addition, parallel algorithms [15] and the multi-rate method [16] are used to reduce computational costs, which performance is affected by the configuration of the power electronics system.

**2)** ***Time constraint***: CHIL simulation typically requires a calculation frequency of 20x-100x the switching frequency to ensure accurate sampling [17], [18]. Therefore, the higher switching frequency of the converters results in a stricter time constraint.

[1]This work was supported by the National Natural Science Foundation of China under Grant 52207209 and U2034201, and funded by China Postdoctoral Science Foundation under Grant 2021M701844. (Corresponding author: *Yangbin Zeng*.)

FPGAs can be used in such small calculation time CHIL simulation time due to their excellent high parallelism, excellent computation speed and low communication latency [19]. FPGA based CHIL simulations with high switching frequencies from 100 kHz to 200 kHz have been implemented in [14], [20]–[22]. Multi-FPGA simulators are developed for scaling up the calculation resources [23]. Besides, [15] proposed a parallel simulation method for the calculation resources optimization which is more suitable for hardware characteristics. High-level synthesis approaches are also used to optimize the performance of FPGA algorithms [24]. In general, increasing the simulation scale requires adding hardware resources or sacrificing simulation accuracy, which is an irreconcilable contradiction under the time constraint.

**3)** ***Simulation accuracy*****:** The power converters of electrified railway use the uncontrolled rectifier bridge to reduce the size of the receiver side of the railway [25]. However, it is difficult to accurately simulate the natural commutation process of the diode by the fixed-step solvers with non-iterative algorithms typically used in CHIL simulations, since the diode switching state is influenced by the external state [26]. Specifically, the chattering phenomenon (high frequency oscillations around the zero crossing point) defined in the [26] occurs very frequently in the diode's discontinuous conduction mode (DCM). The reason is that such solvers have difficulty in finding the exact switch-events of the diode, thus generating chattering. Fig. 2(a) gives as an illustrative example a full-bridge circuit with a diode in the blocked state operating in DCM mode. Fig. 2(b) shows the chattering and delay phenomena of the inductor current for different fixed-step simulation algorithms.

A few studies have noticed this phenomenon and provided solutions, but only on specific topologies or simple topologies to eliminate the chattering phenomenon [26]–[28]. A zero tuning (ZR) method was proposed in [26], which combines the normal operating model function with the zero tuning method to simulate the blocking mode, which is limited by hardware requirements and potential instabilities. In [27], high resistance states were introduced to deal with the natural commutation problem in specific topologies of MMC. Besides, A direct mapping method associates state variables with diode states, avoiding the use of iterative solutions, which is difficult applied to complex topologies [28].

In general, conventional CHIL simulations can be summarized as a time-based synchronization framework (TS-CHIL) that can address some of the challenges mentioned above. A common point of the TS-CHIL is that it has a strict real-time constraint. With this constraint, the TS-CHIL is quite difficult to apply to the railway power converters that contain all the above challenges with limited hardware resources.

In this paper, an event--based synchronization framework for CHIL simulation (ES-CHIL) is proposed from the perspective of breaking the real-time requirement and improving simulation accuracy. The behavior of the real controller and the simulator are synchronized through the event-axis instead of through the time-axis. Thus, the ES-CHIL simulation can ease the real-time constraint and broaden the upper bound on the system size and switching frequency. Meanwhile, a variable-step solver with natural commutation detection method is used for the improvement of simulation fidelity. The main contributions of this paper are as follows:

• The proposed ES-CHIL framework does not have strict constraint for the calculation time, which can greatly increase the upper bound on the system size and switching frequency of CHIL simulation.

• The variable-step solver with a switch-event detection method is applied to the ES-CHIL to eliminate the chattering phenomenon in the natural commutation process and improve the simulation fidelity.

• The ES-CHIL framework can provide CHIL testing of power electronic systems of electric railway that are difficult to implement using conventional methods, thus improving the efficiency and safety of their design and evaluation.

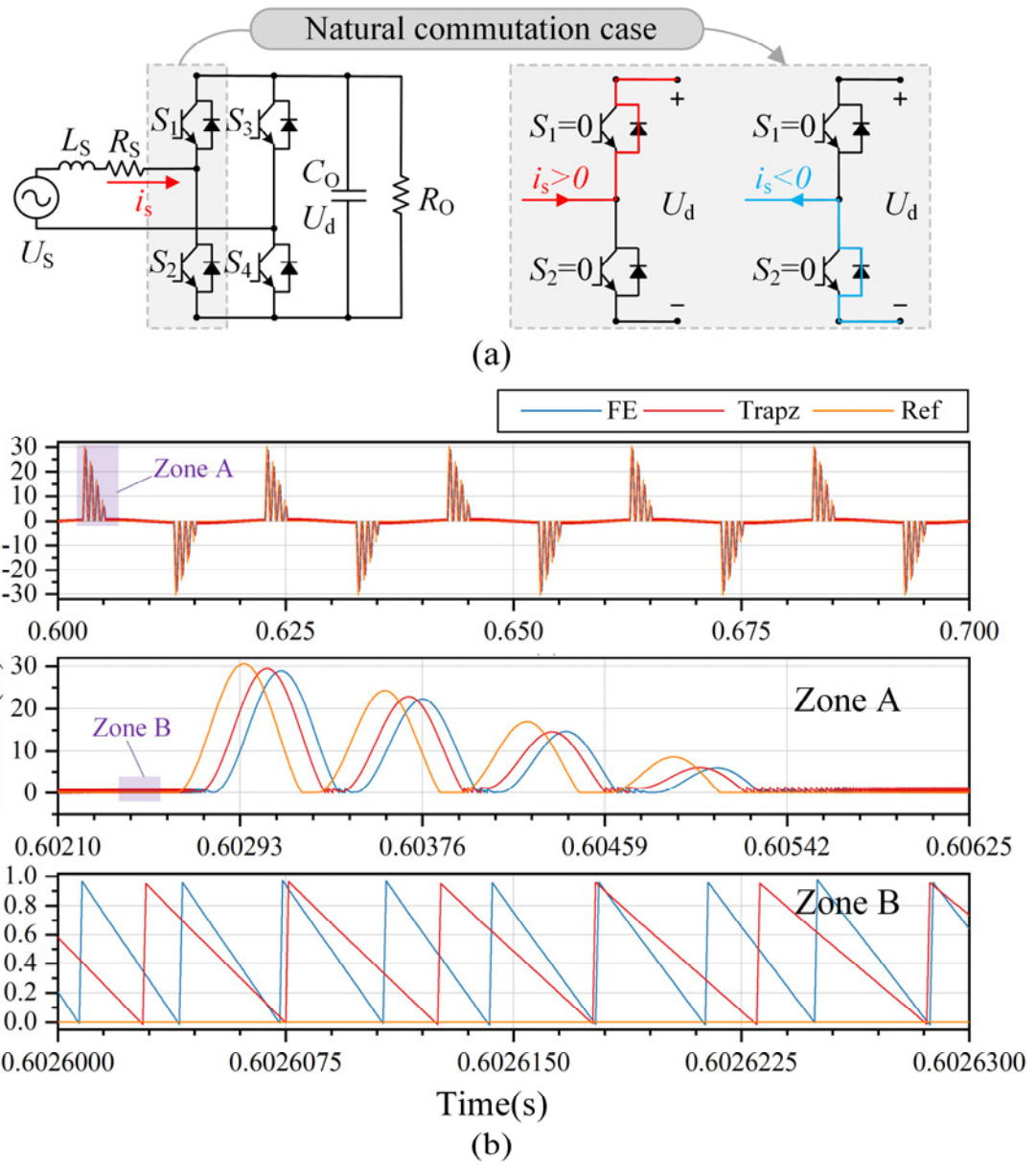

**Fig. 2** Chattering phenomenon during natural commutation. (a). Test circuit in uncontrolled rectification mode. (b). Chattering phenomenon of the current $i_S$ with different fixed-step algorithm.

This paper is structured as follows: the topology and parameters of the studied WPT system is introduced in Section II. The concept of the ES-CHIL is proposed in Section III. In Section IV, the performance of ES-CHIL is evaluated compared with Simulink and field experiments in both time domain and frequency domain. Finally, the conclusion is given in Section V.

## II. The Studied Electric Railway Wireless Power Transfer System

This section introduces the studied WPT system for CHIL simulation, including its topology, control strategy and hardware prototype platform.

### *A. The Topology of the Electric Railway WPT*

The topology of the railway WPT system used for CHIL simulation in this paper is shown in Fig. 3. It consists of three main parts: the transmitter side, the magnetic coupling mechanism and the receiver side, where the transmitter and receiver side adopt a modular structure to increase the power capacity.

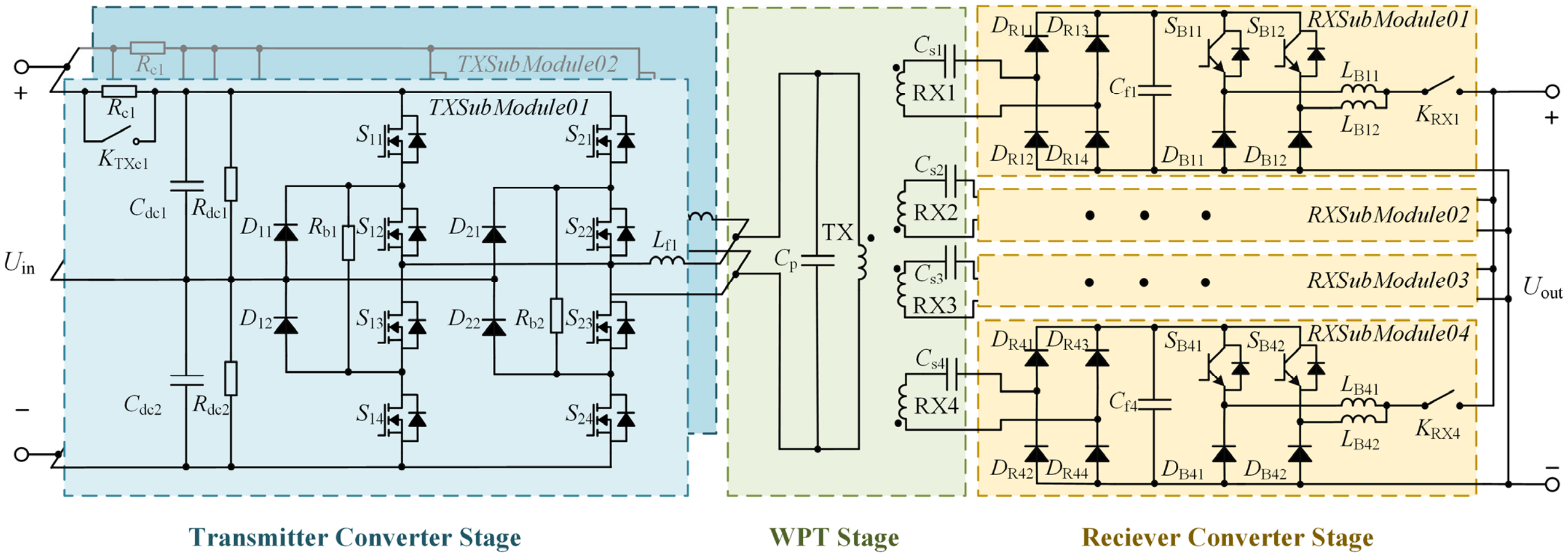

**Fig. 3** The overall circuit of the studied WPT system.

The transmitter side is deployed on the track and the transmitter-side converter consists of two parallel-connected modules (called *TXSubmodule* in Fig. 3). These modules use a diode-clamped three-level full-bridge topology with quasi-three-level modulation. In the *TXSubmodule*$_{01}$ of Fig. 3, $S_{11}$ to $S_{24}$ are full control power switches, $D_{11}$ to $D_{22}$ are clamp diodes, $R_{b1}$ and $R_{b2}$ are equalization resistors, $C_{dc1}$ and $C_{dc2}$ are bus capacitors, $R_{c1}$ is the pre-charge resistor, and $K_{TXc1}$ is the pre-charge bypass contactor.

The magnetic coupling mechanism adopts the LCL-S resonant topology, which has a fixed ratio of output voltage to input voltage. Therefore, the receiver side can be maintained at a stable operating condition at different power levels [29]. In Fig. 3, $L_f$ and $C_p$ are the resonant inductance and capacitance of *TXSubmodule*$_{01}$, respectively.

The receiver is installed on the train and each train contains four receiver modules (called *RXSubmodule* in Fig. 3). Each module uses a diode rectifier bridge to combine light-weighting and power quality. After the rectifier bridge, two parallel Buck circuits are connected to reduce the output ripple by staggering the phase shift. In the *RXSubmodule*$_{01}$ of Fig. 3. $C_{s1}$ is the resonant capacitor, $D_{R11}$ to $D_{R14}$ are the power diodes of the rectifier bridge, $S_{B11}$ to $S_{B12}$ are the full control power switches of the Buck circuit, $D_{B11}$ to $D_{B12}$ are diodes of the Buck circuit, $L_{B11}$ to $L_{B12}$ are the Buck circuit filter inductors, and $K_{RX1}$ is the output contactor.

## B. The Control Strategy of the WPT

The adopted LCL-S resonant topology has the advantage that the current at the transmitter side is not affected by the receiver side [29]. Thus, each side can be controlled independently, reducing the control complexity of the overall system. The control strategies for the transmitter and receiver sides are shown in Fig. 4.

Specifically, the control strategy at the transmitter side is shown in Fig. 4.a. The three-level H-bridge converter in

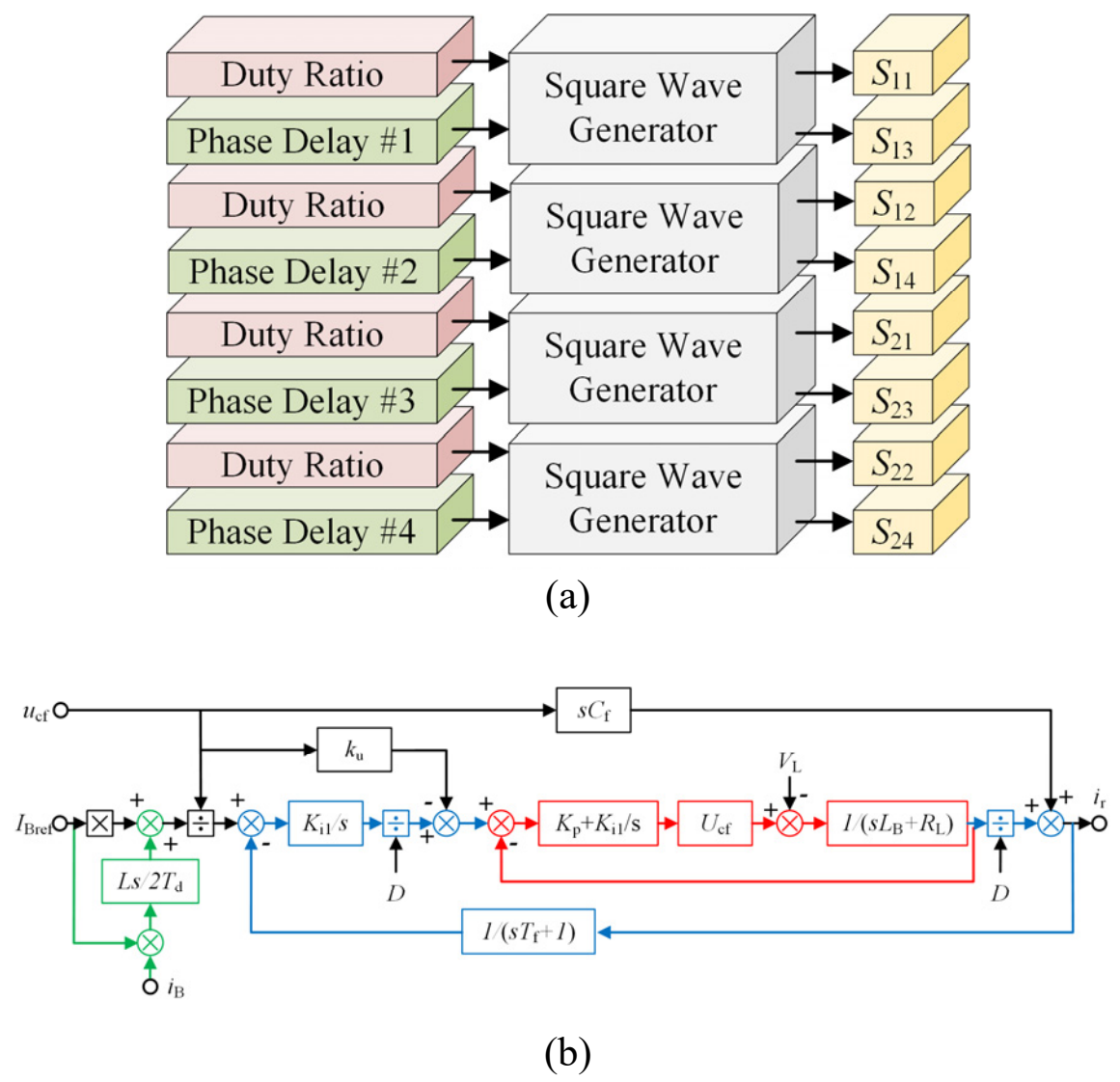

**Fig. 4** Control strategy for the studied WPT system. (a). Transmitter control strategy. (b). Receiver control strategy.

*TXSubmodule* has a switching frequency of $f_0$ = 40 kHz and operates in quasi-three-level mode, i.e., the output voltage of the bridge arm is approximately a square wave with 50% duty cycle, with zero level output only at dead time. The converter starts by phase shifting and thereafter maintained at full phase during normal operation, without adjusting the internal shift.

The Buck control strategy at the receiver side is shown in Fig. 4.b. It controls the output current and reduces the output current ripple by interleaved phase shifting. A modified PI closed-loop control strategy that improves the small-signal impedance characteristics of the above system is used to improve the dynamic response speed [30].

## C. Hardware Prototype Platform and its CHIL Demand

Based on the above topology and control strategy, a 350 kW WPT power converter prototype for electric railway was developed in previous work [25], as shown in Fig. 5. The circuit parameters are shown in Table A-I in Appendix.

The high-power level and specific operating conditions testing of the studied WPT prototype has requires strict requirements for test environment, while the safety risks of prototype testing also need to be considered. Therefore, the CHIL simulation is attempted to verify the system performance. However, since it is difficult to apply the conventional CHIL to the above WPT converter, an ES-CHIL framework is proposed in this paper.

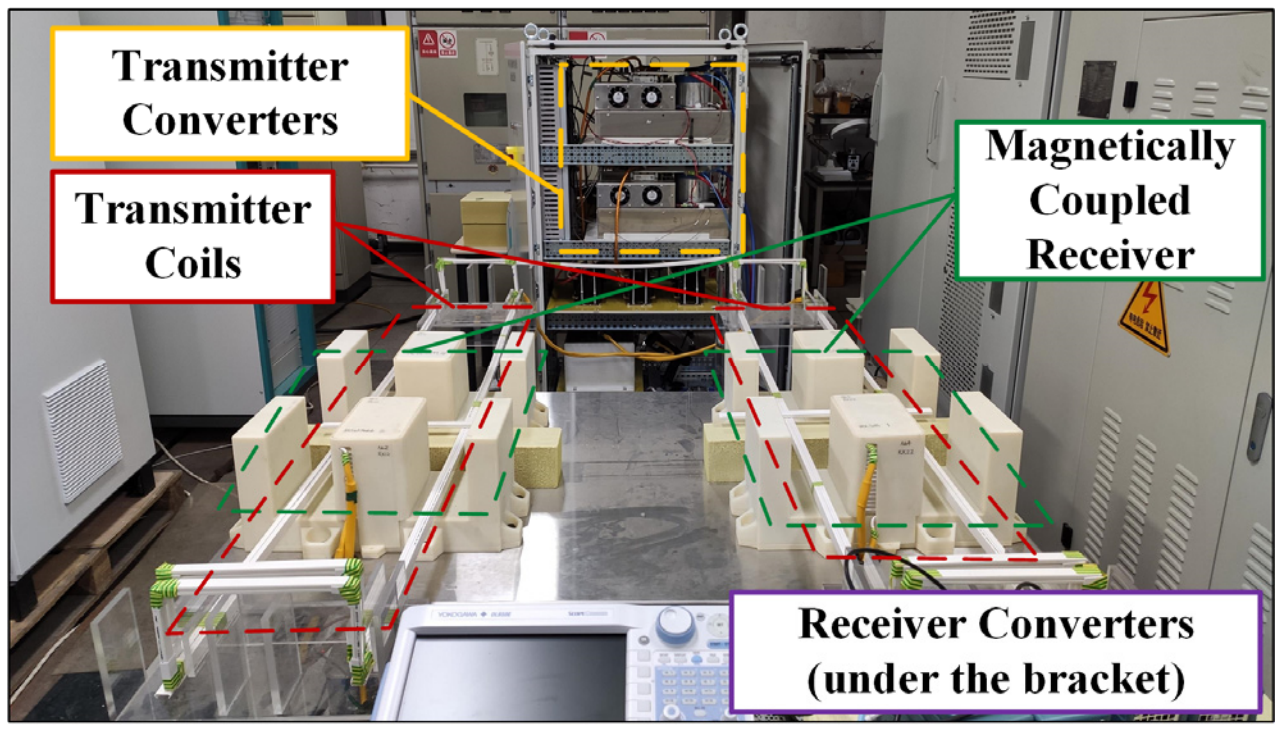

**Fig. 5** The studied WPT prototype.

## III. Event-Based Synchronization Framework

In this section, an event-based synchronization framework is proposed to help implement CHIL simulation function for power electronics systems that cannot implement time-sync-based CHIL simulation.

### A. Multi-layers Event Concept

CHIL system consists of two main parts: the physical digital controller and the power electronic simulator. The power electronic simulator is designed to calculate continuous state values of power electronic systems, and represent the system as a set of ordinary differential algebraic equations (ODEs). Eq. 1 gives the standard form of the ODE model, where $\boldsymbol{x}_s$ refers to the state variables of the system, $\boldsymbol{u}_s$ refers to the input vector, $\boldsymbol{y}_s$ refers to the output vector and $\boldsymbol{t}$ refers to the simulation time. $\boldsymbol{f}_S()$ and $\boldsymbol{g}_S()$ represents the state equation and the output equation, respectively.

$$\begin{cases} \dot{x}_S = f_S(x_S, u_S, t) \\ y_S = g_S(x_S, u_S, t) \end{cases} \quad (1)$$

The physical digital controller consists of sampling, processor and PWM generator, etc. The controller can usually be considered as a discrete-event system that completes control logic operations triggered by a clock signal. The control logic operation can be simplified as

$$y_C[k] = f_C(x_C[k]), k = 1, 2, 3, \ldots, n \quad (2)$$

where $\boldsymbol{x}_c[\boldsymbol{k}]$ and $\boldsymbol{y}_C[\boldsymbol{k}]$ denote the input variable and the output variable of the control system, respectively. $\boldsymbol{f}_C(\ )$ represents the associated signal transmission and information processing.

It can be found that the operational state of the CHIL system is determined by a combination of continuous states from the simulator and discrete events from the controller. Therefore, the whole CHIL system containing discrete events and continuous states is taken as the object of study in this paper, rather than the separate simulator as in the conventional CHIL. Further, the concept of four layers of events in the CHIL system is proposed, and the relationship between them is shown in Fig. 6.

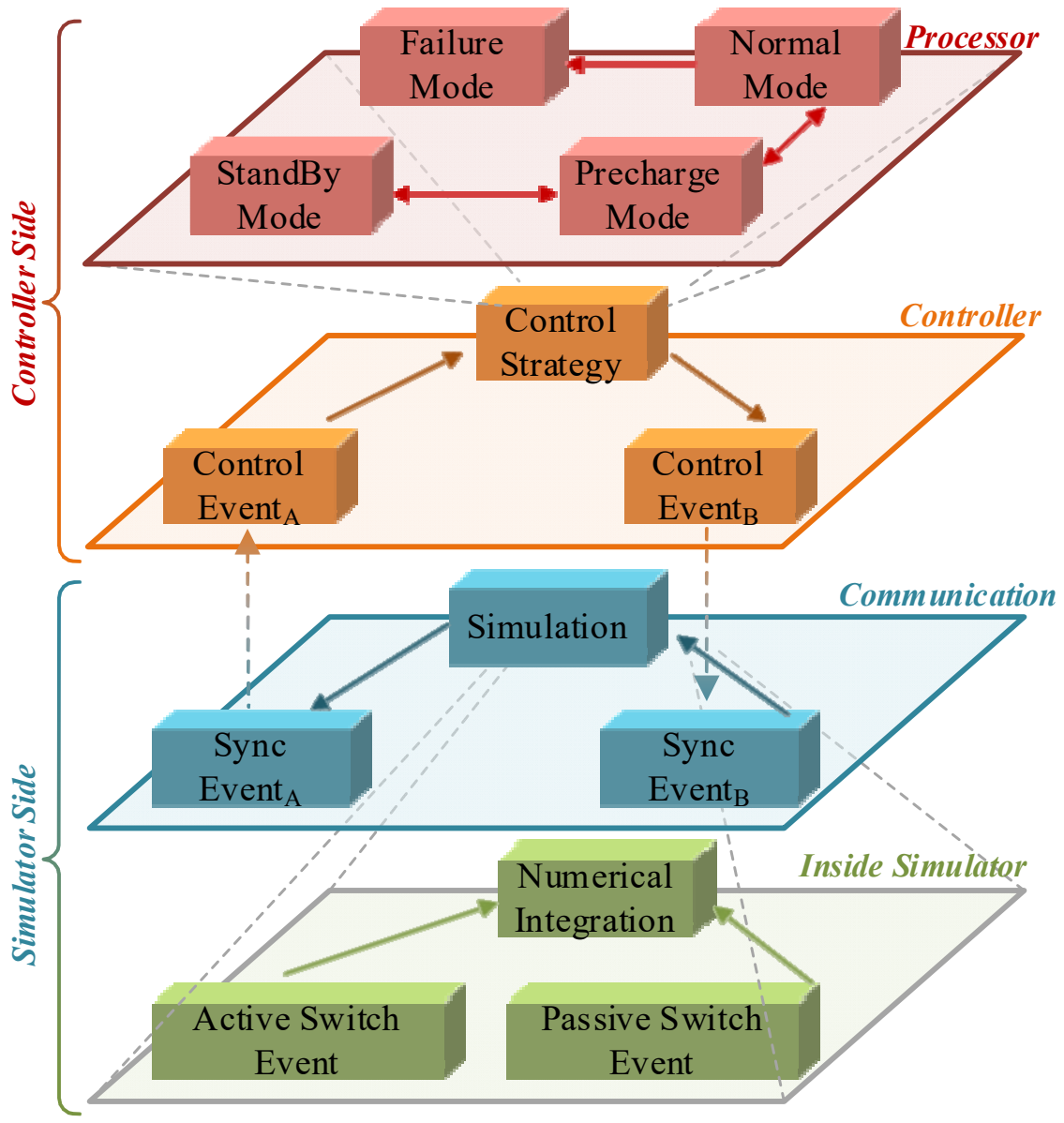

**Fig. 6** Four layers event definition for CHIL simulation.

**1. *Clock-event***: The microprocessor in the control system generally works in the interrupt mode. The behavior of the microcontroller's timed interrupt is defined as a clock-event. The microcontroller executes the control strategy code((include wait mode, pre-charge mode, normal operating mode and fault mode, etc.)) only when a clock-event occurs.

**2. *Control-event***: The control events of the microcontroller consist of two parts. The behavior that the voltage and current sampling values are acquired by the microcontroller is defined as control-event-A ($CE_A$). The behavior that the microcontroller updates the PWM signal to the PWM generator is defined as control-event-B ($CE_B$).

**3. *Sync-event***: The sync-event contains two parts in the communication between the real controller and the simulator: The behavior of the simulator sending voltage and current value to the controller is defined as sync-event-A ($ES_A$), and the behavior of the controller sending control information to the simulator is defined as sync-event-B ($ES_B$).

**4. *Switch-event***: Simulated power systems generally consist of two kinds of switch-event. The behavior of the control signal of controllable power semiconductor changing is defined as active-switch-event (*ASE*). The behavior of the switch state of diode changing is defined as passive-switch-event (*PSE*).

### B. ES-CHIL Framework

The key issue of the CHIL simulation framework is the accurate synchronization between the controller and the simulator. The synchronization diagram of the TS-CHIL (Fig. 7.a) and the proposed ES-CHIL (Fig. 7.b) are given in Fig. 7, respectively.

**1)** ***TS-CHIL*****:** The behavior and the events of the controller are first analyzed, as is shown in Fig. 7.(a). The behavior of the controller during one switching cycle $T_C$ is as follows. First, $C_{1\text{-A}}$ represents the action that the ADC starts to convert the data and is driven by the clock event $p_1$, $t_d$ represents the time interval between the start of ADC sampling and the completion of sampling. After the time $t_d$, the control-event $CE_{1\text{-A}}$ occurs when the microprocessor receives the sampled data. The microprocessor then executes the control strategy code in the period of $t_{run}$. Event $CE_{1\text{-B}}$ represents that the control command signal is sent down to the PWM generator. Driven by the clock signal $p_2$, $C_{1\text{-B}}$ represents the PWM generator generating a new PWM signal according to the control command and sending it to the hardware. The controller enters a new switching cycle at the occurrence of clock event $p_2$.

As for the simulator, the conventional CHIL simulation focuses on the performance inside the simulator, so it is difficult to coordinate the collaboration between the simulator and the controller. Therefore, a fixed time-step synchronization is needed to collaborate with the controller, which is called as TS-CHIL, as shown in Fig. 7(a). The controller and simulator in TS-CHIL run in parallel and are synchronized at predefined time synchronization points ($TS_1$-$TS_6$). The time synchronization points are predetermined, so the TS-CHIL simulators have strict real-time requirements, i.e., $t_{sim} < t_{real}$, where $t_{sim}$ denotes the computation time and $t_{real}$ denotes the time synchronization time. It is necessary to introduce high-speed computing devices and simplify the model to ensure this requirement, which will increase the hardware cost of the simulation and reduce the accuracy of the model [12], [20].

2) ***ES-CHIL*****:** The ES-CHIL simulation framework consists of a simulator, a microcontroller (DSP, ARM, or other floating-point or fixed-point microprocessor) and an event scheduler for linking the two parts.

The event scheduling algorithm of ES-CHIL is shown in Fig. 8. Specifically, the event scheduler determines the sync-event based on the clock-event and the simulation completion signals of the previous cycle. Furthermore, the controller starts the next cycle based on the sync-events generated by the event scheduler, rather than on clock events as in TS-CHIL. Due to the need to monitor the clock events of the controller and the simulation completion signals from the simulator, the event scheduler requires more steps to complete the data exchange than the TS-CHIL.

Since the synchronization is performed by sync-events, ES-CHIL does not need to have a constraint on the computation time as TS-CHIL does. Therefore, the computation time of a switching cycle $Tc$ may be smaller or larger than $Tc$ in ES-CHIL. When the calculation time is less than the switching period in Fig. 7.(b), it means that the calculation completion signal ($SE_1$) occurs before the clock-event $p_1$. Sync-events $ES_{1\text{-A}}$ & $ES_{0\text{-B}}$ occur simultaneously with the clock-event $p_1$, and realize the data exchange between controller and simulator. The Sync-events ($ES_{1\text{-A}}$ & $ES_{0\text{-B}}$) is equivalent to the time-syce point $TS_1$ of TS-CHIL in Fig. 7.(a).

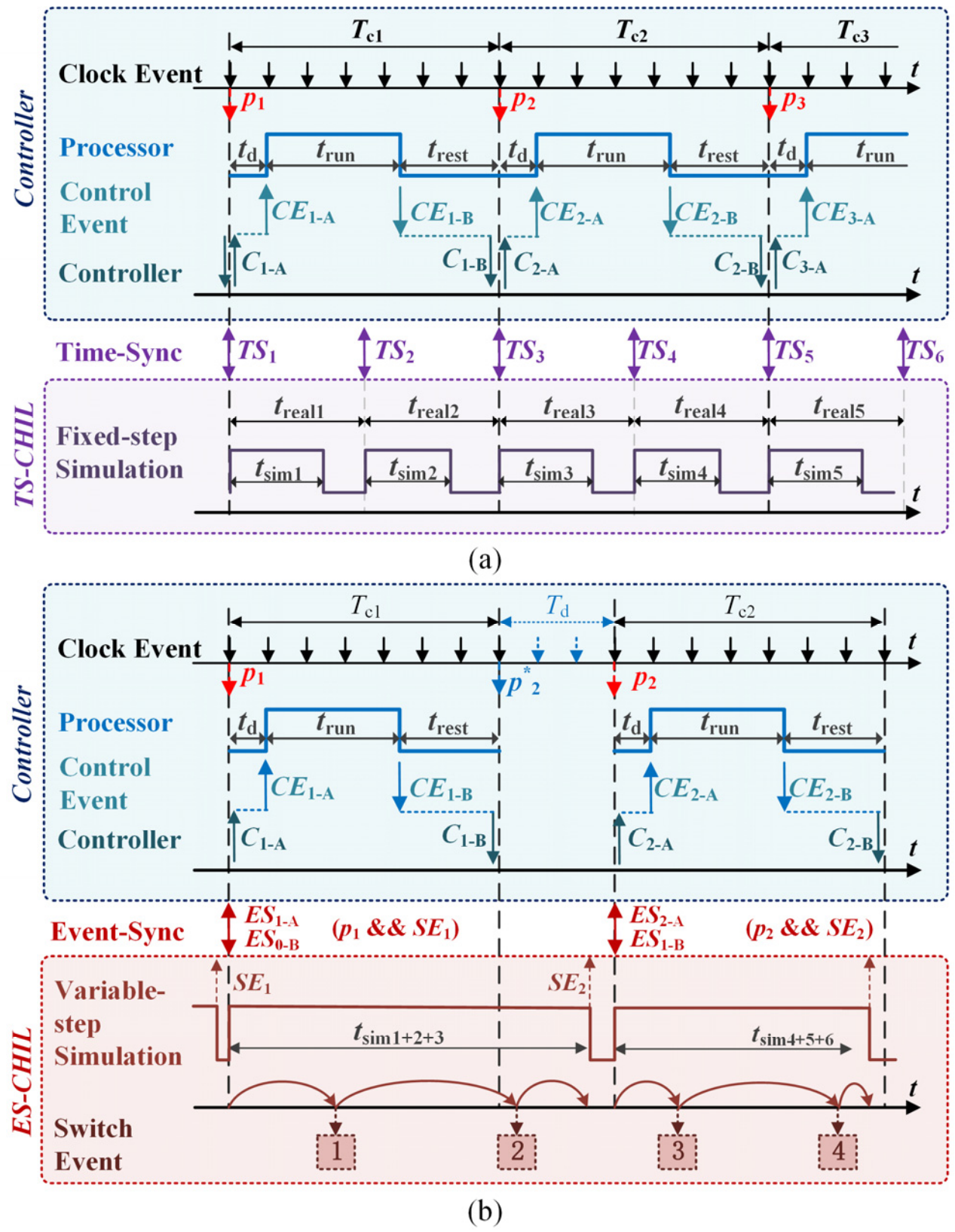

**Fig. 7** Time diagram for CHIL Simulation. (a). TS-CHIL. (b). ES-CHIL.

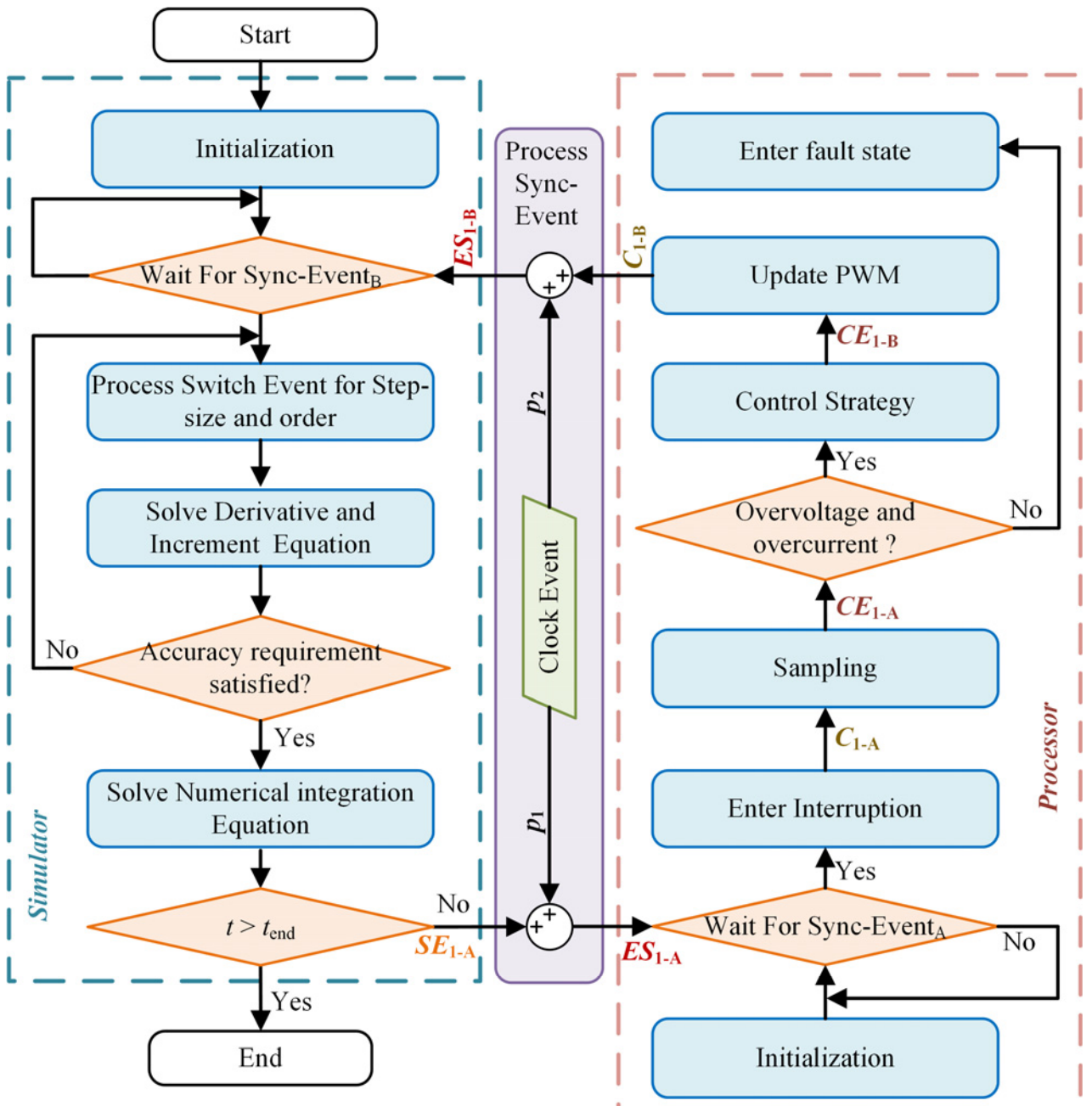

**Fig. 8** Flowchart for the event scheduler in the ES-CHIL framework

When the calculation time $t_{sim1+2+3}$ greater than $T_c$ in Fig. 7.b, it means that the simulation completion signal ($SE_2$) occurs after the clock event $p_2^*$. Therefore, the occurrence conditions for sync-events $ES_{2\text{-A}}$ & $ES_{1\text{-B}}$ are not satisfied when the clock-event $p_2^*$ occurs. The controller enters a frozen state under the action of the event scheduler. The frozen state corresponds to the blue clock event between $p_2^*$ and $p_2$ in Fig. 7.b. $SE_2$ occurs as the simulator completes the computation. Sync-events $ES_{2\text{-A}}$ & $ES_{1\text{-B}}$ occur with the next clock event $p_2$, and the controller is unfrozen and continues its operation.

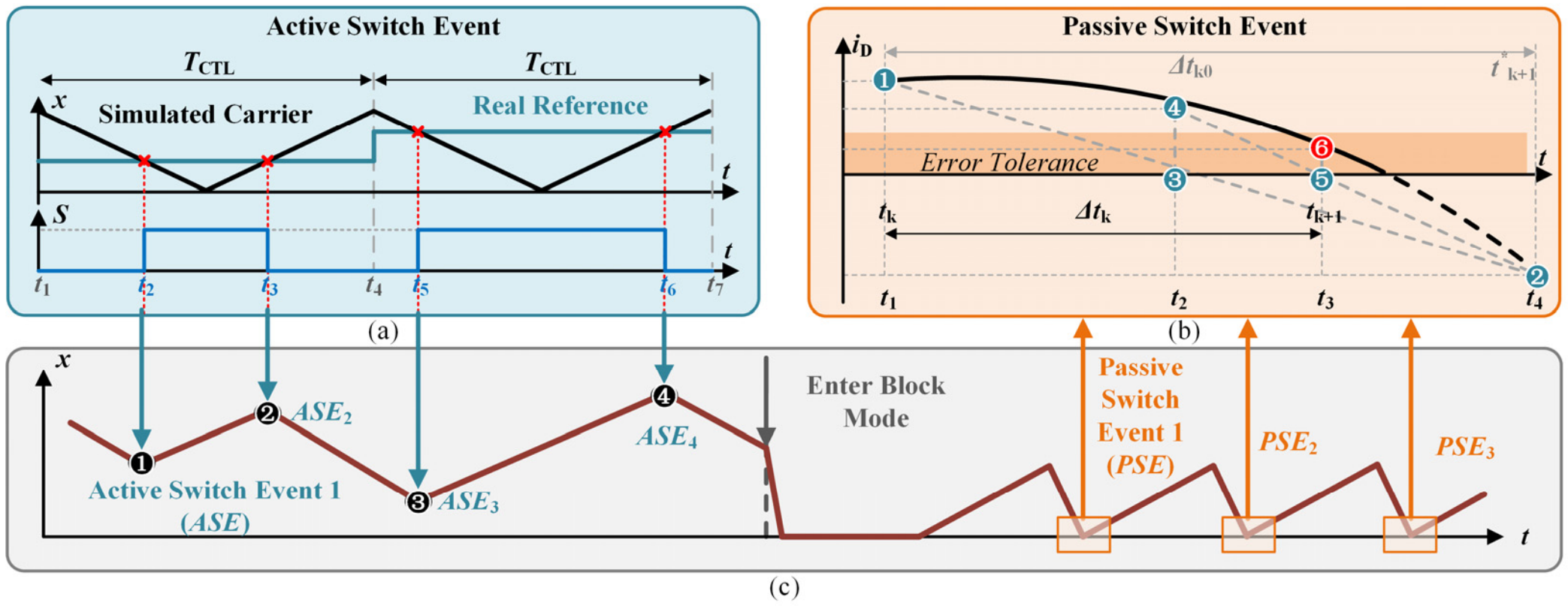

**Fig. 9** Switch event detection method for both active switch event and passive switch event

One advantage of ES-CHIL is that it does not need to comply with strict real-time requirements, allows the use of more detailed models and cheaper computing hardware, etc. Furthermore, the causality of each switching cycle of the controller is preserved due to the monotonically increasing simulation clock. Therefore, the controller in the ES-CHIL framework has the same causality and single-switching cycle behavior as in the real world. Even though the simulator is not running in real-time, the synchronization between the controller and the simulator will not be in an error state and remains the same as the physical world through the proposed ES-CHIL framework.

### C. Switch Event Detection Method

Since no strict time constraint in the ES-CHIL framework, the switch-event detection method with a variable-step solver is proposed to eliminate the chattering phenomenon in the natural commutation process and improve the simulation fidelity.

Specifically, the continuous state equation of Eq. (1) is firstly linearized,

$$\begin{aligned}\dot{\mathbf{x}}(t_k) &= A_k\mathbf{x}(t_k)+B_k\mathbf{u}(t_k)\\ y(t_k) &= C_k\mathbf{x}(t_k)+D_k\mathbf{u}(t_k)\end{aligned} \tag{3}$$

where $\boldsymbol{x}(t)$ represents the independent state variable, consisting of capacitor voltage and inductor current, $\boldsymbol{u}(t)$ represents the independent input variable, and $\boldsymbol{y}(t)$ represents an independent output variable. $\boldsymbol{A}_k$, $\boldsymbol{B}_k$, $\boldsymbol{C}_k$, and $\boldsymbol{D}_k$ map independent state variables and independent input variables to $\boldsymbol{x}(t)$ and $\boldsymbol{y}(t)$, respectively.

Since the coefficients $\boldsymbol{A}_k$, $\boldsymbol{B}_k$, $\boldsymbol{C}_k$, and $\boldsymbol{D}_k$ in Eq. (3) are constants, and $\boldsymbol{x}(t)$ and $\boldsymbol{u}(t)$ can be considered linear in $[t_k, t_{k+1}]$, matching the characteristics of the LTI system. Therefore, the recursive property of the LTI system differential equations can be used to obtain the various derivatives of the state variables and the output variables.

$$\mathbf{x}^{(i+1)}(t_k) = A_k\mathbf{x}^{(i)}(t_k)+B_k\mathbf{u}^{(i)}(t_k), i=0,1,2...n \tag{4}$$

$$\mathbf{y}^{(i)}(t_k) = C_k\mathbf{x}^{(i)}(t_k)+D_k\mathbf{u}^{(i)}(t_k), i=0,1,2...n \tag{5}$$

where $\boldsymbol{x}^{(i)}$ represents the $i$-th derivative of the state variable $\boldsymbol{x}$ and $\boldsymbol{y}^{(i)}$ represents the $i$-th derivative of the output variable $\boldsymbol{y}$.

Therefore, the numerical solution of the state variable at the time of $t_{k+1}$ can be obtained according to the Eq. (4) by using the Taylor series.

$$\begin{aligned}x(t_{k+1}) &= x(t_k)+\sum_{i=1}^{p}\frac{x^{(i)}(t_k)}{i!}\Delta t_k^i + O(\Delta t_k^{p+1})\\ &= \tilde{x}(t_{k+1})+O(\Delta t_k^{p+1})\\ &\approx \tilde{x}(t_{k+1})\end{aligned} \tag{6}$$

where $\tilde{x}(t_{k+1})$ represents the numerical solution of $x(t_{k+1})$, $p$ represents the order of the Taylor series, and $O(\Delta t_k^{p+1})$ represents the local truncation error from the numerical solution to the true solution. The local truncation error is determined by the step size and order of the Taylor series. The step size and order adjustment is explained in detail [31].

In this paper, the switch-events of power semiconductors are divided into two types, as shown in Fig. 9. The switch-events of the fully controlled device before blocking are active switch-events, and the switch-events that occur in the diode after blocking are passive events. The processing of active switch events and passive switch events is expanded below.

Active switch-events (ASE) are generated by the controller's PWM generator. The simulator obtains the simulated PWM signal by comparing the reference wave of the virtual PWM generator with the carrier information of the control event in the real controller. Therefore, the occurrence time and state changes of all active switch-events in a control cycle are known in advance, and thereby the step size can be flexibly arranged to avoid iterative positioning and improve the calculation efficiency, as shown in Fig. 9.a.

The switch-event of diodes is only related to the corresponding system state and is hard to be predicted in advance, so it is called a passive switch-event (PSE). Generally, it is necessary to rely on iterative positioning to find it accurately. The computational cost of the iterative switch-event detection methods is quite hard for time-sync-based CHIL simulator but can be supported in ES-CHIL simulation due to no calculation time limitation.

The details of the iterative switch-event detection method are as follows: passive switch-events need to monitor the corresponding state variable changes to obtain the exact switch-event occurrence time. For example, the moment when the diode current drops to the threshold current (typically 0A) is monitored during the diode turn-off process, and the moment when the diode voltage exceeds the threshold voltage (typically 0.7V) is monitored during the diode turn-on process.

The derivative of each order of the state quantity $\boldsymbol{y}_{\text{diode}}$ to be monitored can be obtained by Eq. (5). Moreover, the derivatives of state variables and input variables have already been calculated in Eq. (6) and do not need to be recalculated. Therefore, judging the switching state of the diode only needs to monitor whether the output variable $\boldsymbol{y}_{\text{diode}}$ reaches the threshold,

$$\underbrace{y_{\text{diode}}(t_{\text{k}})+\sum_{i=1}^{p}\frac{y_{\text{diode}}^{(i)}(t_{\text{k}})}{i!}\Delta t_{k}^{i}}_{y_{\text{diode}}(t_{\text{k+1}})}-\delta_{\text{th}}=0 \tag{7}$$

This paper uses the secant method to solve the root of Eq. (7) to locate the moment when the passive switch-event occurs. Fig. 9.b presents a secant diagram for monitoring the current state during diode turn-off process.

Overall, the concept of multi-layer events for CHIL simulation is presented in this section. Then, an event-sync-based CHIL simulation framework is proposed to solve the drawback of time-sync-based CHIL simulation. Finally, the switch-event-driven simulation is embedded in ES-CHIL framework to improve the accuracy of switch event detection, especially for the passive switch-events. In addition, since there is no time constraint, the transient switching model can be used on the proposed framework to obtain higher fidelity.

## IV. Results Analysis and Performance Evaluation

The hardware platform of ES-CHIL simulation is introduced firstly. Then, the results of ES-CHIL are compared with Simulink® simulation results and experimental results, and the performance of ES-CHIL is evaluated through time and frequency domain analysis.

### A. ES-CHIL Hardware Platform

ES-CHIL does not require high computing hardware capacity. Therefore, a personal computer (PC) is used as the computing hardware to reduce the hardware cost. The hardware platform is shown in Fig. 10. The power circuit of the WPT is deployed on the personal computer, and the control strategy code is deployed on the device under test (DUT) controller, which is connected to the simulator via the PCIe bus.

### B. Accuracy Validation Compared with Simulink®

The ES-CHIL simulation accuracy is verified with Simulink® SimPowerSystems tool, which results serve as the benchmark for analyzing. The control system is implemented in Simulink® and ES-CHIL, respectively. The simulated working conditions are as follows.

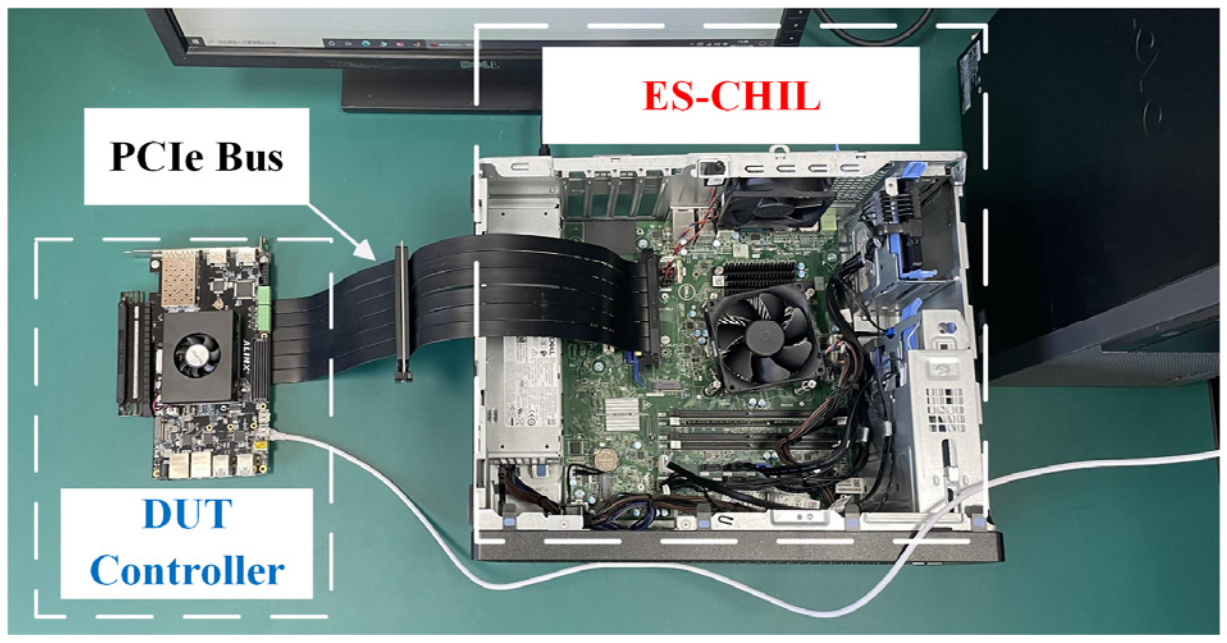

**Fig. 10** Simulation Hardware platform

The transmitter starts by phase shifting within 0.01s and shift phase is not adjusted after starting condition. The receiver does not work when the transmitter is phase-shifted, and starts to work after 0.01s, using closed-loop strategy to control the average value of the output current.

Fig. 11 (a) shows the simulation results of Simulink®, Fig. 11 (b) shows the simulation results of ES-HIL, and Fig. 11 (c) shows the enlarged view of the comparison between the simulation results of Simulink® and ES-HIL. Among them, Irx1 represents the receiving inductor current, Ibuck1 represents the buck output current, and $U_{\text{out}}$ represents the supercapacitor charging voltage.

Typical values of the simulation results are compared in Table I. Relative error of steady-state results below 2%. It shows that the simulation results of ES-CHIL and Simulink® are very close, even in dynamic processes. These subtle differences are due to the inconsistencies in the implementation method of the control system in ES-CHIL and Simulink®.

TABLE I

TYPICAL VALUES OF THE SIMULATION RESULTS

| | ES-CHIL | Simulink | Relative Error |
|---|---|---|---|
| $I_{\text{TX}}$ RMS value /A | 138.5 | 136.2 | 1.69% |
| $I_{\text{rx1}}$ RMS value /A | 93.44 | 94.93 | 1.57% |
| $\underline{U}_{\text{ctx}}$ RMS value /V | 2658 | 2690 | 1.19% |
| $U_{\text{C}}$ RMS value /V | 959.9 | 942.2 | 1.88% |
| $I_{\text{buck11}}$ RMS value /V | 49.32 | 49.96 | 1.28% |
| $I_{\text{out}}$ RMS value /A | 399.9 | 399.7 | 0.05% |

*Measurement time range of steady state value: 0.048s~0.049s
**1) $I_{\text{TX}}$ represents the transmitting coil current; 2) $I_{\text{rx1}}$ represents the receiving coil current; 3) $\underline{U}_{\text{ctx}}$ represents the bus capacitor voltage of the buck circuit at the transmitting side; 4) $U_{\text{C}}$ represents the bus capacitor voltage of the buck circuit; 5) $I_{\text{buck11}}$ represents Buck1 output current; 6) $I_{\text{out}}$ represents the supercapacitor charging current

### C. Controller Behavior Test Compared with Experiment

The integrity of the controller behavior in ES-CHIL is verified by comparing the experimental results. The controller in the ES-CHIL uses the same control code and strategy as the controller in experiment.

The power-loop condition is adopted in experiment and the hardware-in-the-loop simulation for verification. The power loop condition is specified by connecting the receiver output back to the transmitter DC bus, so that the grid only needs to provide a small amount of system power loss.

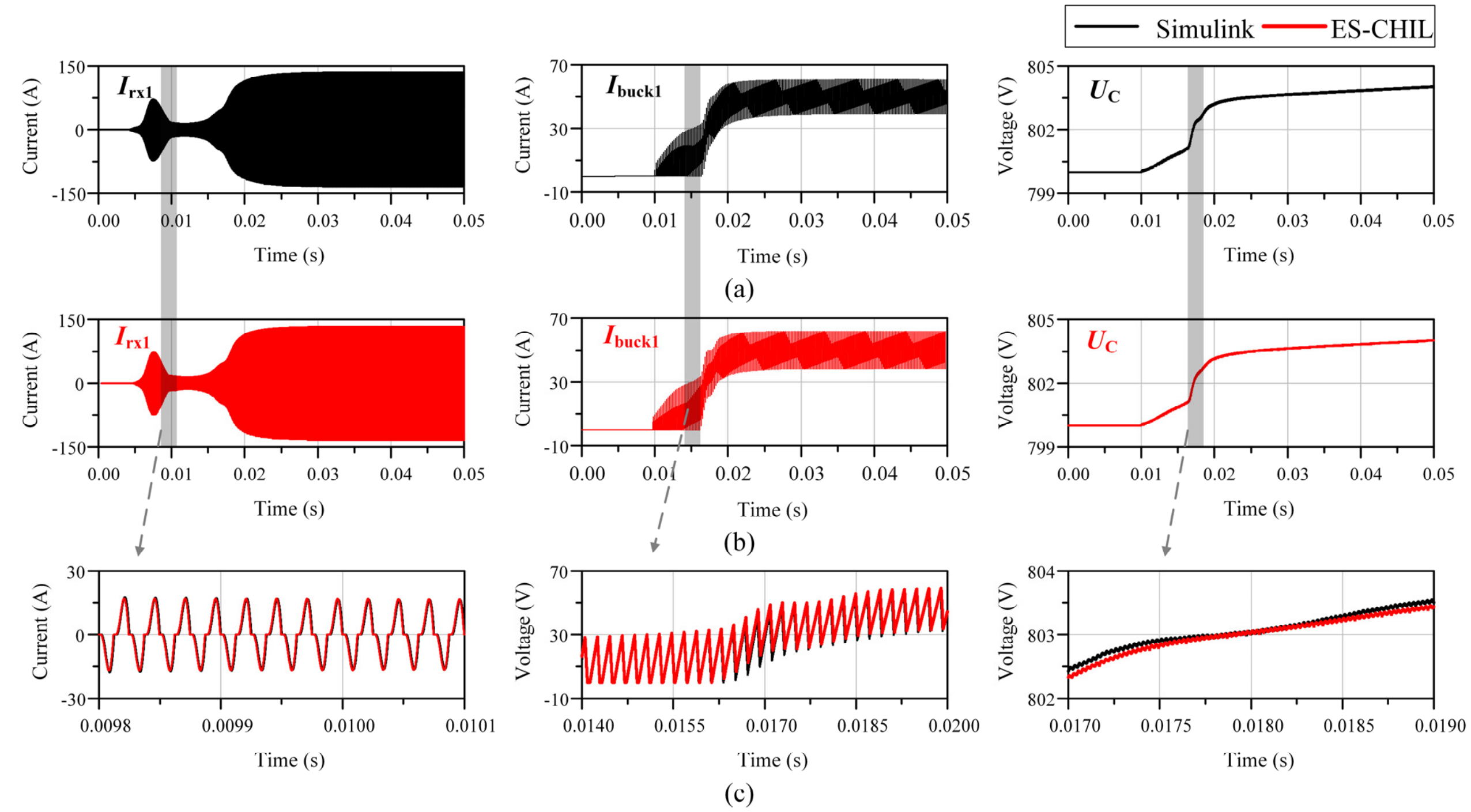

**Fig. 11** Simulated results of the start process of the WPT. (a). The simulation results of Simulink®. (b) The simulation results of ES-CHIL (c). The enlarged view of the comparison between the simulation results of Simulink® and ES-CHIL

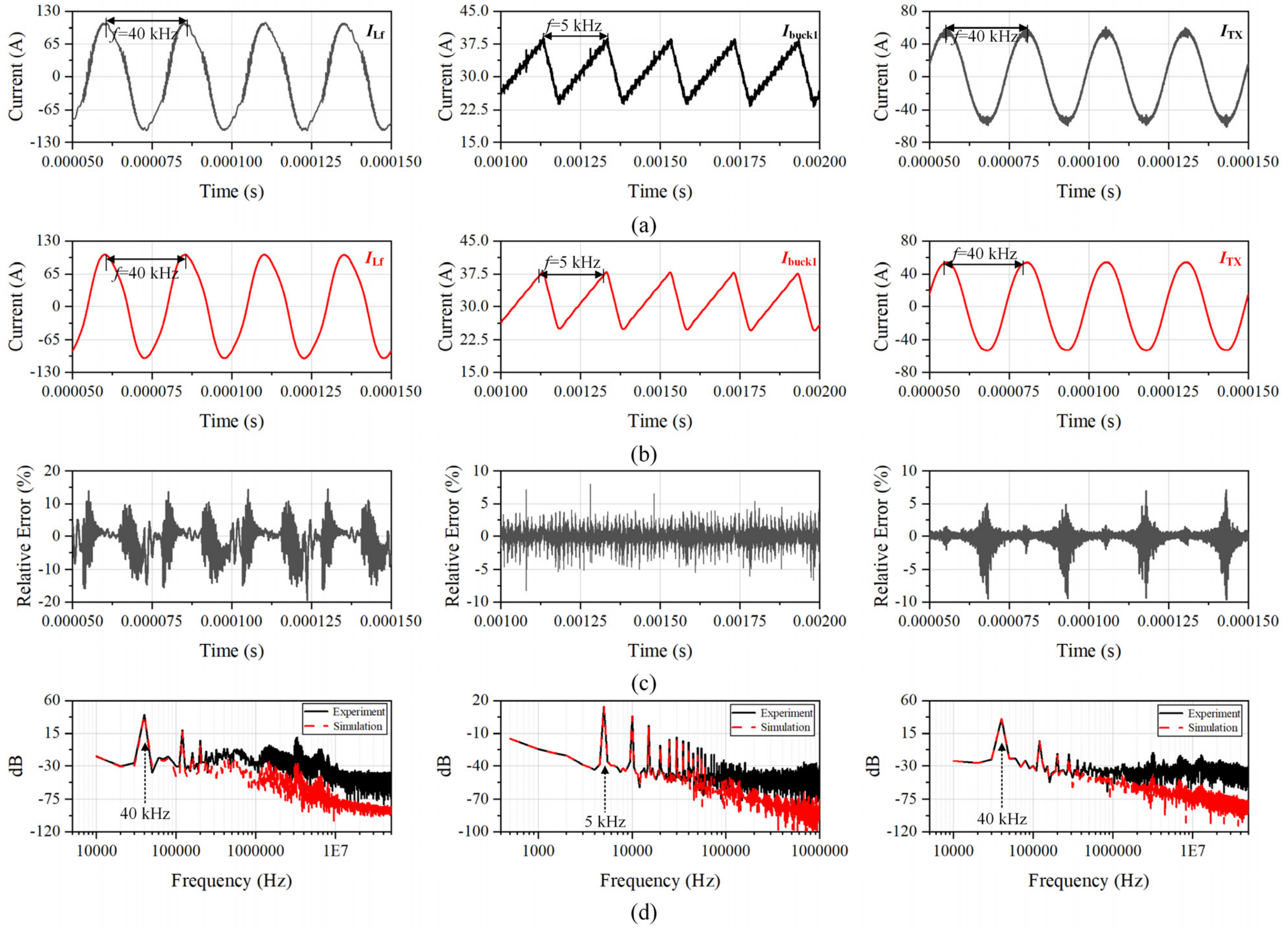

**Fig. 12** Comparison between the ES-CHIL simulation and experiment for the WPT system. (a). Experiment waveform. (b) Simulation waveform (c). Relative error between the ES-CHIL simulation and experiment. (d). FFT for the ES-CHIL simulation and experiment

Fig. 12 (a) shows the experimental waveform, Fig. 12 (b) shows the ES-CHIL simulation waveform, Fig. 12 (c) shows the error comparison between the experimental waveform and ES-CHIL simulation waveform, and Fig. 12 (d) shows the FFT analysis comparison between the experimental waveform and ES-CHIL. Among them, $I_{\text{lf1}}$ denotes the inductor current at of *TXSubmodule*$_1$, $I_{\text{buck11}}$ denotes the output current, and ITX denotes the transmitter coil current.

The better agreement between the waveforms of ES-CHIL and the experiments can be seen from Fig. 12.(a)-(b). The relative errors of the ES-CHIL simulation and experimental waveforms can be seen in Fig. 12.(c). $I_{\text{lf}}$ and $I_{\text{tx1}}$ have large relative errors in some areas due to sampling errors and other elements at the over-zero and highest points, but the overall relative errors are small. Ibuck11 has a small relative error of about 5% because it is a straight flow without over-zero points.

To further illustrate that ES-CHIL does not affect the behavioral integrity of the controller, FFT analysis of the experimental and simulation results are made in Fig. 12.(d), respectively. From the results, the ES-HIL simulation results retain the fundamental frequency information and low frequency information of the experimental results intact, which can restore the waveform of the real device well. The high-frequency information cannot be fully consistent with the experiment due to the sampling error of the oscilloscope of the physical experiment and the ideal modeling used in the simulation. Therefore, ES-CHIL can better play the role of testing the controller program.

### D. Performance Compared with Time-sync-based simulator

In this part, the advantages of the proposed ES-CHIL are illustrated by comparing the proposed ES-CHIL with the time-sync-based CHIL simulator.

The Forward Euler (FE) and Backward Euler (BE) algorithm is widely used the time-sync-based CHIL simulation due to its natural parallelism and small single-step computation. However, in the existing work, step sizes of 100 ns and below have been used only in applications with simple circuits, and the simulated solvers have been limited to low-order algorithms such as FE and BE and their improvements, as detailed information can be found in the Table II. To compare the performance, only one receiver module of the receiver's WPT system is simulated in both the ES-HIL and the TS-HIL. Different steps and different solvers are used in the TS-HIL simulator, and the ode45 solver in Simulink serves as the benchmark. The comparison of the simulation results is shown in Fig. 13 and 14. The advantages of ES-HIL are mainly threefold:

**1)** ***Higher simulation accuracy:*** The simulation accuracy of the TS-CHIL is limited by the real-time constraint. It gives the AC-side and DC-side current results of the uncontrolled rectifier bridge in *RXSubmodule*$_1$, as shown in Fig. 13.(a) and Fig. 13.(b), respectively. Fig. 13.(c) and Fig. 13.(d) are enlarged views of Fig. 13.(a) and Fig. 13.(b), respectively. It can be seen that the resonant current waveform obtained when using 100ns step-size and 50ns step-size simulation has obvious sawtooth, and the simulation results obtained obviously do not have sufficient accuracy, and 10ns or even smaller steps must be used for simulation. Such 10ns step size is also very difficult for FPGA-based real-time hardware-in-the-loop simulators. On the contrary, the ES-CHIL simulation proposed in this paper can achieve the same results as the benchmark results by consuming more time in exchange for higher accuracy because there is no real-time constraint. It indicates that the relative error of the TS-CHIL with 100ns step-size is about 38 times higher than it of the ES-CHIL, and more detailed comparisons are shown in Table III.

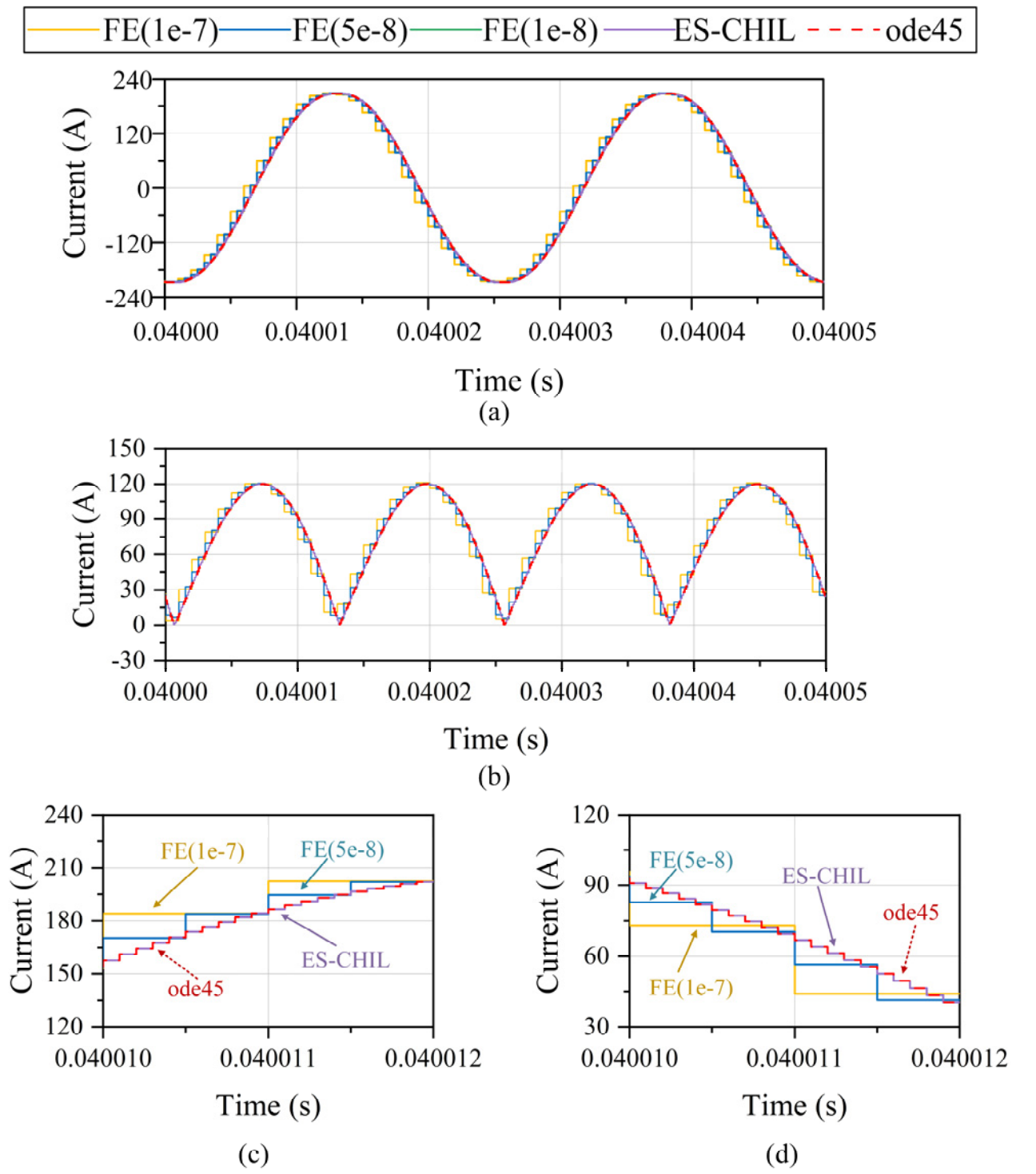

**Fig. 13** The simulation fidelity comparison between ES-CHIL and TS-CHIL. (a). The AC-side current results of uncontrolled rectifier bridge in RXSubmodule1. (b). The DC-side current results of uncontrolled rectifier bridge in RXSubmodule1. (c). Enlarged view of AC-side current results of uncontrolled rectifier bridge in RXSubmodule1. (c). Enlarged view of DC-side current results of uncontrolled rectifier bridge in RXSubmodule1.

TABLE II
COMPARISON WITH THE EXISTING WORK ON THE TS-CHIL SIMULATION

| Reference | Step-size(ns) | Solver | Switching Frequency (kHz) | Application |
|---|---|---|---|---|
| [3] | 15 | FE | 160 | LLC with 8 Sw[1] |
| [29] | 36 | BE | 50 | AC-DC-AC with 12 Sw |
| [30] | 100 | FE | 60 | LLC with 8 Sw |
| [31] | 40 | FE+BE | 40 | NPC with 14 Sw |
| [32] | 40 | FE | 100 | AC-DC with 6 Sw |
| [13] | 250 | TP[2] | 10 | SST with 20 Sw |
| [24] | 25 | BE | 50 | LLC with 8 Sw |
| ES-CHIL | variable | TF[3] | 40 | WPT with 56 Sw |

[1]Swiching Device (including fully controlled device and diode)
[2]Trapezoidal rule
[3] Taylor's Formula Solver.

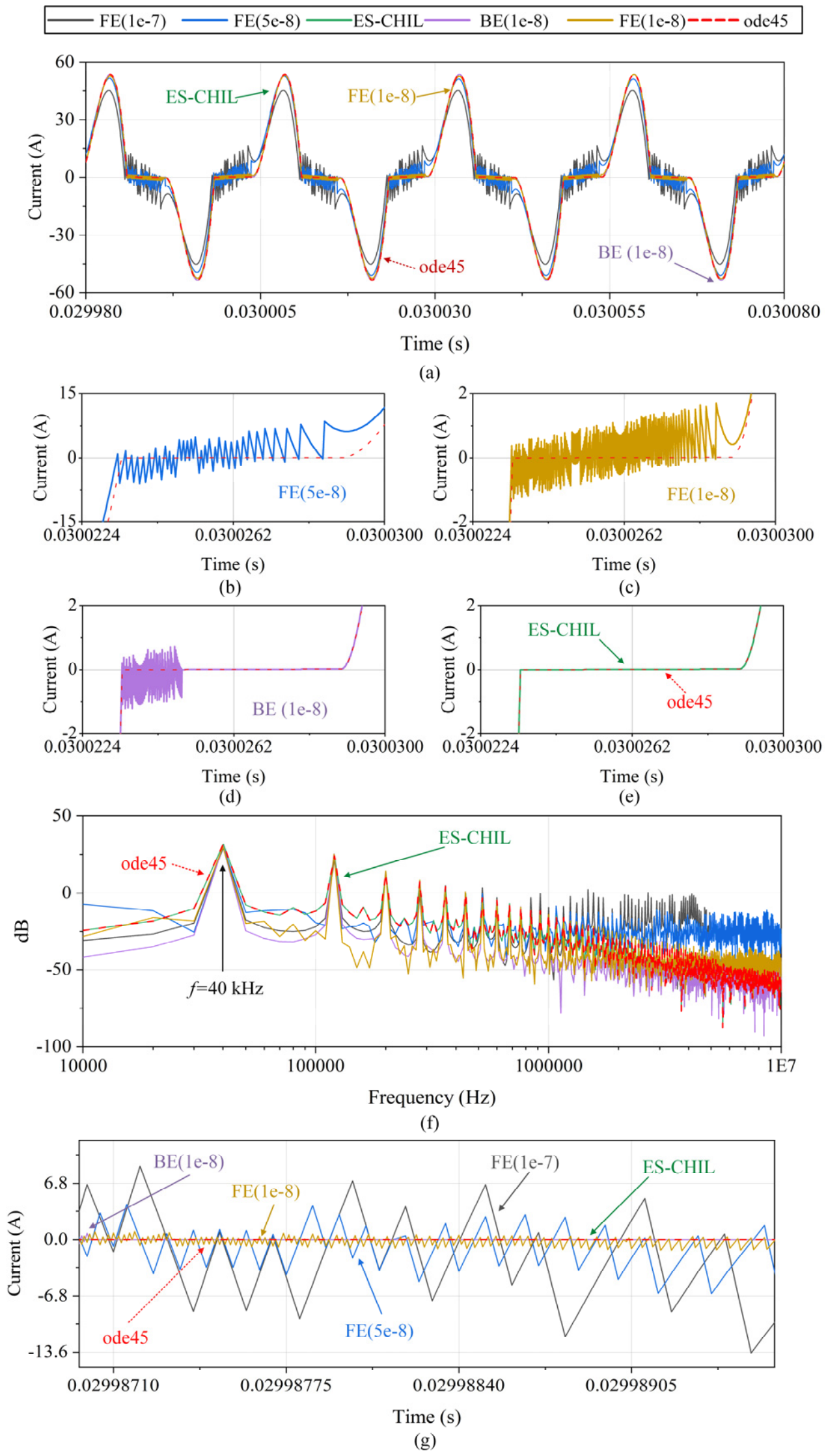

**Fig. 14** Comparison of chattering phenomena generated by different CHIL emulators. (a). The AC-side current results of uncontrolled rectifier bridge in RXSubmodule1 with DCM mode. (b)-(e). Enlarged view of chattering phenomena generated by different CHIL emulators. (f). FFT analysis and comparison. (g). Enlarged view of comparison of chattering phenomena.

**2)** ***No spurious chattering phenomenon:*** When the receiver side of WPT is lightly loaded, the uncontrolled rectifier bridge at the receiver side of the diode may enter the natural commutation process. The TS-CHIL simulator is hard to calculate iteratively. Therefore, it is not possible to find the exact natural commutation point, which will generate chattering near the natural commutation point, as shown in Fig. 14. The chattering at the natural commutation points decreases with the step size. The chattering phenomenon introduces non-characteristic harmonics, which affects the accuracy of the simulation. In this paper, ES-CHIL removes the shackles of real time by the framework of event synchronization, and the exact natural commutation moment can be found by iterative

TABLE III
THE RELATIVE ERROR OF DIFFERENT SIMULATION FRAMEWORK

| Simulation Framework | Solver | Step-size | Relative Error* | Ratio |
|---|---|---|---|---|
| ES-CHIL | TF Solver with SEDM | Variable | 0.0176% | 1:1 |
| TS-CHIL | FE | 100 ns | 0.6675% | 37.82 : 1 |
| | | 50 ns | 0.3392% | 19.22: 1 |
| | | 10 ns | 0.1372% | 7.76 : 1 |

* The compared benchmark is the Simulink results with ode45 solver

TABLE IV
THE ABSULOTE ERROR OF CURRENT IN DCM MODE

| Time | FE (1e-7) | FE (5e-8) | FE (1e-8) | BE (1e-8) | ES-CHIL |
|---|---|---|---|---|---|
| 0.029988 | 7.1248 | 3.0017 | 0.5755 | -0.0017 | -0.0018 |
| 0.030012 | 6.6396 | -2.4004 | 0.8425 | 0.5058 | 0.0026 |
| 0.0300244 | 4.2611 | 2.6297 | 0.7897 | 0.5374 | -0.0027 |
| 0.0300369 | -4.2609 | -2.4606 | 0.6306 | 0.0021 | 0.0011 |
| 0.0300744 | 4.2606 | -5.0846 | -0.6705 | -0.9853 | 0.0061 |
| Average | 5.3094 | 3.1154 | 0.7017 | 0.4065 | 0.0029 |
| Radio | **1830:1** | **1074:1** | **242:1** | **140:1** | **1:1** |

TABLE V
COMPARISON IN HARDWARE COST

| Reference | Platform | LUTs | DSP48 | BRAM | Price |
|---|---|---|---|---|---|
| [11] | Kintex® - XC7K410T | 45560 (17.9%) | 43 (2.8%) | 106 (13.3%) | $3,019.0 |
| [30] | Virtex® - XC7V485T | 116670 (38%) | 762 (27%) | 1202 (3%) | $5,244.0 |
| [31] | Kintex® - XC7K410T | 47028 (25%) | 120 (17.7%) | 91 (11.6%) | $3,019.9 |
| [32] | Zynq® - XC7Z020 | Not mentioned | | | $1,074.0 |
| [8] | NI® PXIe-7975R | 123611 (48.6%) | 644 (41.8%) | 91 (11.4%) | $17499.0 |
| [25] | Kintex® - XC7K325T | 1095 (0.5%) | 88 (1.05%) | 22 (4.9%) | $2,544.0 |
| ES-CHIL | PC | CPU: intel® i7-11700<br>RAM:4.8Gb/16 GB<br>OS: Ubuntu® 18.04 | | | **$323.0** |

search. The average errors of ES-CHIL and different time-sync-based solvers with different step-size are given in Table IV. It can be seen that the average error of the time-sync-based solver even with 10ns step size is still 140 times higher than the error of the ES-CHIL solver.

**3)** ***Low Hardware Cost:*** Compared with other TS-CHIL simulators, the ES-CHIL does not need high-performance computing hardware and therefore has low hardware costs. time-synchronous-based simulators rely on high-speed computing hardware, such as FPGAs and Zynqs, and the simulation scale is related to the hardware resources. This will definitely increase the hardware cost for larger simulation sizes. Compared to other time-synchronization-based simulators, ES-CHIL does not rely on high-speed computing hardware and therefore has low hardware costs. The hardware cost comparison is shown in Table V. time-synchronous-based simulators rely on high-speed computing hardware, such as FPGAs and ARMs, and the simulation size is related to the hardware resources. For larger simulation scales, this definitely increases the hardware cost. Besides, the Intel CPU used in ES-HIL has much lower hardware cost than other TS-simulators, which facilitates the promotion.

### *E. Limitation and Future work*

The limitation of the ES-CHIL is hard to test the whole controller, mainly for the processor in the controller, including testing the control code and control logic, etc. The A/D sampling, PWM generation and other functions are hard to be tested. Nevertheless, ES-CHIL is helpful for the control pre-testing of the converter prototype design and develop process, due to its convenient way of establishment and low cost.

The proposed ES-CHIL can ease the real-time constraint and broaden the upper bound on the system size and switching frequency. Therefore, it is possible to adopt different simulation solvers in the ES-CHIL framework, or even to coordinate multiple simulation solvers for co-simulation using the proposed ES-CHIL simulation framework in the future.

## V. Conclusion

An ES-CHIL simulation framework is proposed in this paper, which synchronized the controller and simulator via the event axis rather than the time axis. It allows enough time to improve the fidelity of the modeling and algorithms while ensuring the behavioral integrity of the controller's single control cycle. Besides, there are no restrictions on the topology and switching frequency of the object to be simulated. Thus, the problem of difficult high-fidelity CHIL simulation of complex topology high-switching-frequency power converters in electrified railway systems has been solved. The proposed ES-CHIL is applied to a 350-kW wireless charging system with a switching frequency of 40 kHz and 56 power switches, and the advantages of the proposed ES-CHIL are verified compared with offline simulation, TS-CHIL and field experiments. The results show that the proposed ES-CHIL has a significant advantage over TS-CHIL in terms of the supported system size and switching frequency, and allows the use of models and algorithms with higher fidelity. Therefore, the proposed ES-CHIL can meet the CHIL requirements for power converters for electrified railway with large system scale and high switching frequency. Further, ES-CHIL can be used in the design and evaluation of these power converters, accelerating the transition to rail electrification and achieving net zero goals faster.

## Appendix

TABLE A-I
SYSTEM PARAMETERS OF THE 350 kW WPT

| Transmitter side | Value | Receiver side | Value |
|---|---|---|---|
| $U_{in}$ | 1500 V | $U_{out_init}$ | 800 V |
| $f_{NPC}$ | 40 kHz | $f_{buck}$ | 5 kHz |
| $L_p$* | 39.5 μH | $L_{si}$* | 255 μH |
| $C_p$ | 442.8 nF | $C_{si}$ | 62.34 nF |
| $R_p$* | 70 mΩ | $R_{si}$* | 150 mΩ |
| $M_{psi}$* | 25.4 μH | | |
| $C_{dc1}$, $C_{dc2}$ | 2750 μF | $C_{f1}$, $C_{f2}$, $C_{f3}$, $C_{f4}$ | 400 μF |
| $R_{dc1}$,$R_{dc2}$ | 47 kΩ | $U_{out}$ | 900 V |
| $L_{fi}$ | 42.95 μH | $L_{B11}$, $L_{B12}$ | 1.1mH |
| $R_{b1}$,$R_{b2}$ | 300 kΩ | | |

*where $L_p$, $L_{si}$, $R_p$, $R_{si}$ and $M_{psi}$ are the equivalent resistance and inductance of the magnetically coupled mechanism.